\documentclass[aps,groupedaddress,nofootinbib,showpacs,english,nobalancelastpage,floatfix,superscriptaddress,preprintnumbers, prd]{revtex4}
\usepackage[latin1]{inputenc}
\pdfoutput=1
\usepackage{multirow, amsmath, amssymb, amsfonts, array}
\usepackage{color}
\usepackage{bm,graphicx,cancel}
\usepackage{calc}
\usepackage{dcolumn}
\usepackage{mathrsfs}
\usepackage[bookmarks=false]{hyperref}

\newcommand{\gammasp}{\gamma_{sp}}

\newcommand{\be}{\begin{equation}\begin{array}}
\newcommand{\ee}{\end{array}\end{equation}}

\begin{document}

\title{Diffuse Gamma Ray Background from Annihilating Dark Matter in Density Spikes around Supermassive Black Holes}

\author{Alexander Belikov}
\affiliation{Institut d'Astrophysique de Paris, UMR 7095 CNRS, University Pierre et Marie Curie, 98 bis boulevard Arago,
75014 Paris, France}
\author{Joseph Silk}
\affiliation{Institut d'Astrophysique de Paris, UMR 7095 CNRS, University Pierre et Marie Curie, 98 bis boulevard Arago,
75014 Paris, France}
\affiliation{1 Beecroft Institute of Particle Astrophysics and Cosmology, Department of Physics,
University of Oxford, Denys Wilkinson Building, 1 Keble Road, Oxford OX1 3RH, UK}
\affiliation{The Johns Hopkins University, Department of Physics and Astronomy,
3400 N. Charles Street, Baltimore, Maryland 21218, USA}

\date{\today}

\begin{abstract}
Dark matter annihilation is proportional to the square of the density and is especially efficient in places of highest concentration of dark matter, such as dark matter spikes.
The spikes are formed as a result of contraction of the dark matter density profile caused by adiabatic growth of a supermassive black hole at the center of the dark matter halo or subhalo.
We revisit the relation between the properties and mass functions of dark matter halos and spikes, and propose alternative models for the density spikes that can potentially contribute to the isotropic gamma-ray background.
\end{abstract}

\maketitle

\section{Introduction}

It has been established that dark matter plays a major role in structure formation. It gravitates towards the overdensities left after the epoch of inflation and collapses into halos. Finally, baryons condense at halo centers and form stars and gas that constitute galaxies. There are implications that most galaxies contain supermassive black holes (SMBH) at their centers \cite{Ferrarese:2000se}.
Primordial dark matter density spikes~\cite{Gondolo:1999ef} are formed around central supermassive black holes in galaxies, within the influence sphere of the black hole. The slope of the dark matter spike $d \ln\rho/{\rm d}\ln r$ depends on the initial slope of the hosting halo $\gamma$ and is equal to $\gammasp = \frac{9-2\gamma}{4 - \gamma}$ (hereafter referred to as the standard spike), with a plateau forming at a radius where annihilations become important over the black hole's life-time. 

There are a number of physical mechanisms that might affect the formation of the dark matter spike and reduce the annihilation rate.

First of all, the dark matter density profile changes its central slope to become a spike only when the black hole grows from at most 1/200th of its final mass at the center of the halo. If the black hole grows from a seed of higher mass the density profile in the center is just rescaled, while the slope remains unchanged \cite{Ullio:2001fb}.

A black hole seed formed off-center has to spiral towards the center of the potential well via dynamic friction before it can induce the formation of a dark matter spike. The black holes of smaller masses require greater spiral-in times. 
In case of the Milky Way halo and the Milky Way SMBH the black hole seed would have to have formed within 50 pc of the center of dark matter distribution for  the standard dark matter spike to be formed.

After having formed the dark matter spike might be dampened due to gravitational interaction with the dense stellar core.
This mechanism is efficient in case of dynamically "old" stellar cores such as that of the Milky Way Galaxy ($\sim$5 stellar relaxation times) but is far less important in the case of subhalos where there are no dense stellar cores (the vast majority in the standard CDM model) nor in massive galaxies, which are dynamically young (their central stellar relaxation times are much longer than a Hubble time). The latter case has been studied in \cite{2008PhRvD..78h3506V}.

In the case of the Milky Way Galaxy, the SMBH growth time is comparable to the stellar heating time, and the outcome for the spike profile has not yet been definitively determined. The typical SMBH of interest for our model turns out to be in a mass range comparable to that expected for massive subhalo dwarfs, and any putative stellar component in the cores may well be partially destroyed by early AGN feedback, as possibly has occurred in one plausible resolution of the so-called "too big to fail" problem 
\cite{2010ApJ...725..556S, 2011MNRAS.415L..40B,2012MNRAS.427.2625P}.

Another process that with certainty disrupts the central dark matter spike is halo mergers. If both merging halos contain SMBHs, in the formation of a SMBH binary and its subsequent decay the energy is transferred to dark matter particles thus lowering the density in the inner regions.
With the help of N-body simulations it was shown that the initially steep dark matter density profile $\rho \sim r^{-2.4}$ is diminished to $\rho \sim r^{-0.5}$ \cite{Merritt:2002vj} up to 100 pc radius for a range of ratios of masses of merging halos from 1:1 to 1:10 with stronger disruption of the spike corresponding to ratios of masses closer to unity. 
However such major mergers of galaxies are generally considered  to play a minor role in AGN triggering \cite{2014arXiv1401.5477V} and therefore presumably in growth. 

The simulations of merger history of halos show that a Milky-Way sized halo undergoes at least one merger since redshift 2 with a halo of mass $2\times 10^{11}\rm M_\odot $ or greater.
At the same time the observed stellar cusp at the Milky Way galactic center suggests that this process might have been not effective.
While in the specific case of the Milky Way or even neighboring halos there might be strong arguments against dark matter spikes, no such argument can be applied to the totality of dark matter halos. Such dark matter spikes can contribute to the isotropic gamma-ray background (IGRB) measured by {\em Fermi} LAT \cite{Abdo:2010nz}. 
This possibility has been considered earlier for supermassive \cite{Ahn:2007ty}, as well as intermediate-mass black holes \cite{Horiuchi:2006de}.

The dwarf galaxies that host the SMBHs of interest to us here are much less likely to undergo mergers with other dwarfs. Hence black hole mergers, considered to be one of the most dangerous ways to destroy DM spikes, are probably much less important in affecting spikes than dynamical heating by central star clusters, discussed above.

In this paper we consider alternative definitions of the spike radius for SMBHs, study the relative enhancement of the gamma-ray signal due to the spikes for dark matter particle masses in the TeV range and point out that the contribution of dark matter spikes to the IGRB comes from a narrow range of halo masses, potentially spoiling the isotropic properties of the gamma-ray background flux.

\section{Extragalactic gamma-ray background from dark matter spikes} 
\subsection{Calculation of the flux}

In this section ,we calculate the contribution of dark matter spikes associated with black holes hosted in centers of galaxies to the isotropic gamma ray background. In what follows, $M_{sp}$ refers to the mass of the dark matter spike, formed in the dark matter halo of mass $M$ around a supermassive black hole of mass $M_{bh}$. Masses are given in units of solar mass. The Schwarzschild radius of the black hole is denoted by $r_S(M_{bh}) = \frac{GM_{bh}}{c^2}$.

Following refs. \cite{Bergstrom:1997fj, Horiuchi:2006de}, we can write down the gamma-ray flux (the number of photons per unit area, solid angle, time and energy) as: 
\be{l}
\label{eq:flux}
\frac{d\Phi_\gamma}{dE_0} = \frac{dN_\gamma}{dAd\Omega dt_0 dE_0} = \frac{c}{4\pi} \int dz \frac{e^{-\tau(E_0,z)}}{H_0 h(z)} \int dM_{sp} \frac{dn}{dM_{sp}}(M_{sp}, z) \frac{d\mathcal{N}}{dE}(E_0(1+z),M_{sp},z),
\ee
where $c$ is the speed of light, $H_0 = 67.9 \frac{\text{km/s}}{\text{Mpc}}$ is the Hubble constant and $h(z) = \sqrt{\Omega_\Lambda + \Omega_m(1+z)^3}$ parametrizes the evolution of the Hubble constant, with $\Omega_m = 0.307$ and $\Omega_\Lambda = 0.693$ being the energy densities of dark matter and dark energy correspondingly. Throughout the paper we use the 2013 Planck results for cosmological parameters \cite{Ade:2013zuv}. $\frac{d\mathcal{N}}{dE} (E,M_{sp}, z)$ denotes the differential energy spectrum of photons emitted per unit time from a dark matter spike of mass $M_{sp}$ at redshift $z$ and for the case of annihilating dark matter can be written in the form:
\be{l}
\label{eq:phDensity}
\frac{d\mathcal{N}}{dE}(M_{sp}, z, z_f)
 = \frac{\sigma v}{2M^2_X} \frac{dN_\gamma(E)}{dE} \int
 \rho^2(r) d^3r.
\ee
In the above formula, $\sigma v$ is the annihilation cross-section, $M_X$ is the mass of the dark matter particle, and $\frac{dN_\gamma(E)}{dE}$ is the annihilation spectrum of photons.

The enhancement of the annihilation rate over uniformly distributed dark matter is denoted $\Delta^2(z)$, which allows us to rewrite eq.~(\ref{eq:flux}) as
\be{l}
\frac{d\Phi_\gamma}{dE_0} = \frac{\sigma v}{8 \pi} \frac{c}{H_0} \frac{\bar\rho^2_0}{M^2_X} \int dz \frac{e^{-\tau(z,E)}}{h(z)} \frac{dN}{dE} (E(1+z)) \Delta^2(z),
\ee
where $\bar \rho_0 = \Omega_m \rho_c$, with $\rho_c$ being the critical density. The gamma-ray attenuation is $\tau(z,E)$ is taken from \cite{Stecker:2006eh}.
Our definition of $\Delta^2(z)$ agrees with that given in \cite{Horiuchi:2006de} and can be cast as:
\be{l}
\Delta^2(z) = \int dM_{sp} \frac{dn}{dM_{sp}} (M_{sp}, z) \,\bar\rho^{-2}_0 \int \rho^2(r) d^3r.
\label{eq:deltaSq}
\ee

The mass of the dark matter spike depends on the mass of the central black hole and can be derived using either $M_{bh}-\sigma$ relation or attributing to the spike the volume which integrated mass equals few times the mass of the black hole (see section \ref{sec:SpikeRadius}).

It has been established that nearly every galaxy contains a SMBH at its center \cite{Kormendy:1995er}. The total mass of the galaxy, including the dark matter halo was shown to correlate with the mass of the central SMBH \cite{Ferrarese:2000se}.
On that basis we can write down the mass function of SMBHs in terms of the halo mass function and express the annihilation rate of a dark matter spike associated with a particular halo in terms of the parameters of the halo.

\subsection{Dark matter density contraction around a SMBH}
Dark matter halos studied in N-body simulations are described by spherically symmetric density profiles and can contain either a cusp ($\rho \underset{r\to 0}{\propto} r^{-\gamma}$) or a core ($\rho \underset{r\to 0}{\propto} const$).
Here, we assume a cuspy halo density profile defined as $\rho(r) = \rho_0\left( \frac{r}{a}\right)^{-\gamma}$.
In the vicinity of a black hole the dark matter halo density undergoes adiabatic contraction.
The modified density profile is set by $\rho_{sp}(r) = \rho_1\left( \frac{r}{r_{sp}}\right)^{-\gammasp}$.
In models with a finite core, the exponent is equal to $3/2$, while in models with initial cusps the spike attains $\gamma_{sp} = \frac{9-2\gamma}{4-\gamma}$ \cite{Gondolo:1999ef}.
However, in the general case of non-circular orbits and dynamical heating by a central stellar cusp, the plateau in the central region is substituted by a mild spike $\rho \sim r ^{-1/2-\beta}$ with $\beta > 0$, where $\beta$ is the anisotropy coefficient of particle orbits. The difference in annihilation rates for such a mild central spike vs a central plateau is expected be in the ballpark of 10\% \cite{Vasiliev:2007vh}. Hence in this paper we use only the central plateau model.

By equating the density of the spike to the density of the halo at the radius of the spike $\rho(r_{sp}) = \rho_{sp}(r_{sp})$, we obtain the density parameter of the spike: 
\be{l}
\rho_1 = \rho_0\left( \frac{r_{sp}}{a}\right)^{-\gamma}.
\ee

We emphasize that the relation between the initial slope of dark matter density and the final slope used in this paper is valid for either initially isotropic velocity distribution or for the case of circular orbits. The possible effects of dampening of the dark matter density by a stellar core, black hole merger and off-center formation of the SMBH are neglected in this approach. More generally models with finite initial central density can result in spikes with slopes in the range $1.5 \leq \gamma_{sp} \leq 2.25$ and initially cuspy dark matter profiles can result in spikes with slopes in the range $2.25 \leq \gamma_{sp} \leq 2.5$.

\subsection{The outer radius of the spike}
\label{sec:SpikeRadius}
We consider two definitions for the outer radius of the spike. In the first approach, we define the spike radius as the radius of gravitational influence of the black hole: the radius at which the kinetic energy of dark matter particle is of the order of its potential energy in the vicinity of the black hole $r_{sp} = \frac{2GM}{\sigma^2}$, where $\sigma$ is the orbital velocity. The orbital velocities of dark matter particles can be estimated by the stellar dispersion velocity. In the second approach, we define the radius of the spike as the radius at which the enclosed mass is equal to twice the mass of the black hole $M (r < r_{sp}) = 2 M_{bh}$ \cite{Merritt:2003qc}.
In the first case, the spike radius is proportional to the Schwarzschild radius of the black hole $r_S(M) = 9.7 \times 10^{-6} \left (\frac{M_{bh}}{10^8} \right) \textrm{pc}$. In order to estimate the velocity dispersion of dark matter particles, we use the the $M_{bh}-\sigma$ relation between the masses of supermassive black holes and velocity dispersion of their host bulges \cite{Ferrarese:2000se}.

The $M_{bh}-\sigma$ relation for black holes is cast in the form:
\be{l}
\frac{M_{bh}}{ 10^8 } = A_{M\sigma} \left ( \frac{\sigma}{200 \, \textrm{km}/s} \right )^{\gamma_{M\sigma}}.
\ee
So in the first case, the spike radius $r_{sp}$ depends on the mass of the halo and the parameters of the $M_{bh}-\sigma$ relation :
\be{lll}
r_{sp} &=& 37.9 \, \left (\frac{A_{M\sigma}}{3.1}\right)^{2/\gamma_{M\sigma}} \left (\frac{M_{bh}}{10^{8} } \right)^{1-2/\gamma_{M\sigma}} \textrm{pc}.
\ee
Hereafter we assume $\gamma_{M\sigma} = 4$ and $A_{M\sigma} = 3.1$ for the sake of simplicity. Recent studies show that $\gamma_{M\sigma}$ varies from 4.22 \cite{Volonteri:2011mm} to 5.64 \cite{McConnell:2012hz} depending on the chosen sample and analysis techniques.
In the second case, $r_{sp}$ depends only on the parameters of the halo, including the initial slope $\gamma$:
\be{lll}
r_{sp} = a \left(\frac{3-\gamma}{2\pi} \frac{M_{bh}}{a^3\rho_0}\right)^{1/(3-\gamma)}.
\ee

\subsection{The inner radius of the spike}
\label{sec:InnerRadius}

The change of the density profile of dark matter halo due to dark matter annihilations is usually neglected due to the fact that the depletion time-scale  (proportional to the dark matter density at a given radius) is much greater than the characteristic time of halo growth.

On the contrary, the density of dark matter in dark matter spikes can be so high that this effect should be taken into account.
The dark matter density at the very center is depleted due to annihilations at a rate $\dot n \sim -\langle\sigma v\rangle n^2$ and the number density evolves according to
\be{l}
n_X(r,t) = \frac{n_X(r,t_f)}{1+n_X(r,t_f)\sigma v(t-t_f)},
\ee
where $t$ corresponds to the current epoch $t=t(z)$ and $t_f$ is the time of formation of the black holes (and spikes around them).

Using this equation in the limit $(t-t_f) \gg (n_X(r,t_f)\sigma v)^{-1}$, we find that the critical density at time $t$ is 
\be{l}
\rho_{lim} = \rho_{sp}(r_{lim}) = \frac{M_X}{\sigma v (t - t_f)} .
\ee

The inner radius of the spike is defined as the radius at which the density of the spike is equal to the critical density:
\be{l}
\label{eq:rLim}
r_{lim}(z) = r_{sp}\left(\frac{\rho_{lim}(z)}{\rho_1}\right)^{-1/\gamma_{sp}} = 
r _{\lim}(z=0) 
\left(\frac{M_X}{100 \textrm{GeV}}\right)^{-1/\gamma_{sp}}
\left(\frac{\sigma v}{3\times 10^{-26} \textrm{cm}^3 s^{-1}}\right)^{1/\gamma_{sp}} 
\left(\frac{t(z)-t_f}{t(0) - t_f}\right)^{1/\gamma_{sp}}. 
\ee
The inner radius is bounded from below by $r_c$ equal to few times the Schwarzschild radius of the black hole. In the non-relativistic calculation, it has been found that density of dark matter vanishes below $r_c = 4 r_S (M_{bh})$ \cite{Gondolo:1999ef}, while the relativistic treatment improves this estimate and finds that $r_c = 2 r_S (M_{bh})$ \cite{Sadeghian:2013laa}. 

In order to parametrize the uncertainty in our knowledge of the inner radius of the spike that can be disrupted by the infall of baryonic of matter into the black hole, or by halo mergers~\cite{Merritt:2002vj}, we redefine $r_{lim}$ by scaling it up by a factor $\kappa$, compared to that defined in eq.~(\ref{eq:rLim}): $\tilde r_{lim} = \kappa r_{lim}$, where $\kappa \ge 1$.

For $r < \tilde r_{lim}$ we assume the spike transforms into a plateau of constant density equal to $\tilde\rho_{lim} = \kappa^{-\gammasp} \rho_{lim}$. The plateau extends from $r_c$ to $\tilde r_{lim}(z)$.

An alternative method of weakening the spike consists of rescaling of the normalization of $\rho_{sp} (r)$ rather than increasing the limiting radius $r_{lim}$~\cite{Ahn:2007ty}. 
The normalization rescaling is time-dependent with a characteristic timescale set by $T_{heat}$, proportional to the squre root of the SMBH and inversely proportional to the effective stellar mass and to the logarithm of the number of stars within the spike,
as a result of solving the Fokker-Planck equation for the phase space density function of dark matter particles in the presence of dark matter-star scattering.
Although well-motivated from the point of view of stellar heating of dark matter the derivation of the density normalization rescaling neglects dark matter annihilation \cite{Merritt:2003qk}.

The most general prescription for changing the density profile of the spike according to relevant physical processes would combine changing $r_{lim}$, $\gamma_{sp}$ and $\rho_1$.

\subsection{Annihilations in the spike}

From eq.~(\ref{eq:deltaSq}) we see that the integral of the square of the density $\bar\rho_0^{-2} \int \rho^2(r) d^3r$ over the region of the spike defines the average enhancement $\Delta^2(z)$.
This integration region can be split into the plateau region $r_c < r < \tilde r_{lim}$, where the density is constant $\tilde \rho_{lim} = \rho_{sp} (\tilde r_{lim}) = \rho_1\left( \frac{\tilde r_{lim}}{r_{sp}}\right)^{-\gamma_{sp}}$ and the spike region $\tilde r_{lim} < r < r_{sp}$:
\be{l}
\label{for:itot}
I_{tot}
= \int\limits^{r_{sp}}_{r_c} \rho(r)^2 d^3r = I_{pl} + I_{sp}
\simeq 4\pi \int\limits^{\tilde r_{lim}}_{r_c} \tilde \rho^2_{lim} r^2 dr +
4\pi \int\limits^{\infty}_{\tilde r_{lim}} \rho^{2}_1 r^{2\gamma_{sp}}_{sp}r^{2 - 2\gamma_{sp}} dr 
\simeq \frac{4\pi}{3} \frac{2\gammasp}{2\gammasp - 3} r^3_{sp} \tilde\rho^{2 - 3/\gammasp}_{lim} \rho^{ 3/\gammasp}_1.
\ee
The ratios of the first and second integrals is just $I_{pl}/I_{sp} = (2\gammasp - 3)/3$. Under the change $r_{lim} \to \kappa r_{lim(0)}$, the integral of the square of the density changes accordingly $I_{tot} \to \kappa^{3-2\gamma_{sp}} I_{tot}$.
With $\alpha = 10^{-3}$, $\gammasp = 7/3$ and $z_f = 9$, formula (\ref{for:itot}) is reduced to
\be{l}
\rho^{-1}_0 M^{-1} I_{tot}(M, z, z_f) = 
3.7 \times 10^{11} 
\left ( \frac{ \langle \sigma v \rangle}{3\times 10^{-26} cm^3 s^{-1}} \right )^{-5/7}
\left (\frac{M_X}{100\,\text{GeV}}\right)^{5/7} 
\left (\frac{M}{10^{12} } \right)^{2/7} 
\left(\frac{t(z) - t(z_f)}{t(0) - t(z_f)}\right)^{-5/7}. 
\ee

We note that the concentration $c$ of the hosting halo drops out from the final formula for $I_{tot}$ for both definitions of $r_{sp}$ (based on the $M_{bh}-\sigma$ relation and the one based on the enclosed mass $M (r < r_{sp}) = 2 M_{bh}$), and so the flux from a dark matter spike depends only on the mass of the hosting halo.

\subsection{Dark matter spike mass function}

Since we assume that every halo contains a SMBH and the formation of every SMBH generates a dark matter spike, we may use the halo mass function to estimate the number of spikes.

Both the relation between the mass of the spike and the SMBH defined Sec. \ref{sec:SpikeRadius} and the relation between the SMBH mass and mass of the halo are needed to calculate the annihilation rate from a dark matter spike, corresponding to a halo of mass $M$.

Evidence correlating the masses of supermassive black holes and masses of hosting dark matter halos was found in the form of the $\sigma-\text{v}_{\text{c,obs}}$ relation \cite{Ferrarese:2002ct}, where $\sigma$ is the bulge stellar velocity dispersion and $\text{v}_{\text{c,obs}}$ is the observed circular velocity. $M_{bh}-\sigma$ relation and the estimate of the total mass of the halo based on the circular velocity allow to cast $M-M_{bh}$ relation, generally written as $M_{bh} = \alpha 10^8 (M / 10^{12})^\beta$. The exponent $\beta$ of the power-law has been found to vary from 1.55 to 1.82, depending on the sample of galaxies and on analysis techniques \cite{Ferrarese:2002ct, Bandara:2009sd, Booth:2009zb}. 

This correlation has been later found not to hold for bulgeless galaxies, which also turns out to generally be less massive \cite{Kormendy:2011sz}.
A later study \cite{Volonteri:2011mm} clarified the reasons behind the absence of such a correlation for lower mass galaxies: SMBH in massive halos are formed at high redshifts and their growth is mainly stimulated by dark matter halo mergers, most frequently for the most massive halos, whereas SMBH in the less massive halos are formed at low redshift.

Finding the correspondence between the properties of the spikes and the halos is complicated by the fact that the SMBHs are formed at redshifts $z \simeq 6-10$, while the relation $M_{bh}-M$ is known at redshift $z=0$.

This difficulty can be approached by using the relation $M_{bh} = \alpha 10^8 (M(z=0) / 10^{12})^\beta$ for the halo of mass $M(z=0)$ at redshift zero and then evolving it back to redshift $z_f$ to obtain $M(z=z_f)$ \cite{Ahn:2007ty}, that can be used to estimate the number density of SMBHs at redshift $z_f$ and later through the halo mass function at $M(z=z_f)$ $\frac{dn}{dM(z_f)}(M(z_f), z)$. In this case the integration should carried over $dM(z_f)$.
For the sake of simplicity we consider a linear relation (corresponding to $\beta = 1$) between the SMBH mass and the halo mass at $z_f$, and fix $\alpha = 10^{-3}$. 
The mass function of black holes is then equal to the halo mass function at the formation redshift $\frac{dn_{bh}}{dM_{bh}} (M_{bh}, z) = \frac{dn_{halo}}{dM(z_f)} (M(z_f), z_f)$. 
That simplifies our calculations and allows us to draw qualitative conclusions.

We assume that the mass function of spikes is proportional to the mass function of SMBHs, which in turn is proportional to  the halo mass function given by 
\be{l}
\frac{dn}{dM}(M,z) = \frac{\bar\rho_0}{M} \frac{\ln\sigma^{-1}(M,z)}{dM} f(\sigma^{-1}(M,z)),
\ee

where $\sigma(M,z)$ is the variance of the linear density field and $f(\sigma^{-1})$ is the multiplicity function. 

The variance $\sigma(M,z)$ can be calculated from the the matter power spectrum $P(k)$ 
\begin{equation}
\sigma^2 (M,z) = D^2(z) \int^\infty_0 P(k) W^2(k, M) k^2dk,
\end{equation}
where $W(k, M) = (3/k^3 R^3)[\sin(kR) - kR \cos(kR)]$ is the top-hat function, at radius $R = (3M/4\pi\bar\rho_0)^{1/3}$. The growth function, $D(z)$, is the linear theory growth factor, normalized to unity at $z=0$~\cite{Eke:1996ds}. The power spectrum $P(k) \propto k^{n_s} T^2(k)$, where $n_s = 0.9635$ is the spectral index and the transfer function $T(k)$ for adiabatic CDM is corrected to take into account the baryonic density $\Omega_b$ \cite{Bardeen:1985tr, Sugiyama:1994ed}. 

We use the ellipsoidal (Sheth-Tormen) form of the multiplicity function~\cite{Sheth:1999mn}, 
\begin{equation}
f(\sigma) = A\frac{\delta_{sc}}{\sigma}\left(1 + \left(\frac{\sigma^2}{a\delta^2_{sc}}\right)^p\right)\exp\left(\frac{a\delta^2_{sc}}{\sigma^2}\right),
\end{equation}
where $p = 0.3$, $\delta_{sc} = 1.686$ and $a = 0.75$~\cite{Sheth:2001dp}. We fix $A = 0.32$ by the requirement that all of the mass resides in halos.

In all scenarios we place one SMBH per halo and assume that all SMBHs are formed at the formation redshift $z_f$, i.e. take the SMBH mass function to be equal to the halo mass function at $z_f$, and to remain constant at redshifts $z < z_f$.
We assume that every halo contains one SMBH of mass fixed by the linear $M_{bh}-M$ relation.
The dark matter spike mass is set by either $M_{bh}-\sigma$ relation or by $M (r < r_{sp}) = 2 M_{bh}$, while their mass function is equal to the SMBH mass function of the corresponding mass.

In one of our scenarios, we also test the possibility that the black holes and the spikes evolve with redshift along with the halo, that is we assume
$\frac{dn_{bh}}{dM} (M_{bh}, z) = \frac{dn_{halo}}{dM} (M, z)$. 

The mass function of SMBH is expected to evolve with redshift, due to mass accretion and halo mergers \cite{Tamura:2005kc, Tanaka:2008bv}. These phenomena also affect dark matter spikes. While mass accretion might have comparatively mild and perhaps positive effect, the disruptive effect of halo mergers on dark matter spikes might be dramatic: initially steep dark matter density profile $\rho \propto r^{-2.4}$ was shown to degrade to $\rho \propto r^{-0.5}$ \cite{Merritt:2002vj}. The relation between the mass function of the spikes and the mass function of SMBH is therefore not straightforward. However, by discarding such effects due to their complexity and assuming the mass functions of the spikes and of the SMBHs to be equal we find the estimate from above on the number of dark matter spikes. 

\section{Analysis}

\begin{figure*}[h]
\centering
\begin{tabular}{ccc}
\includegraphics[width=0.45\linewidth,clip=]{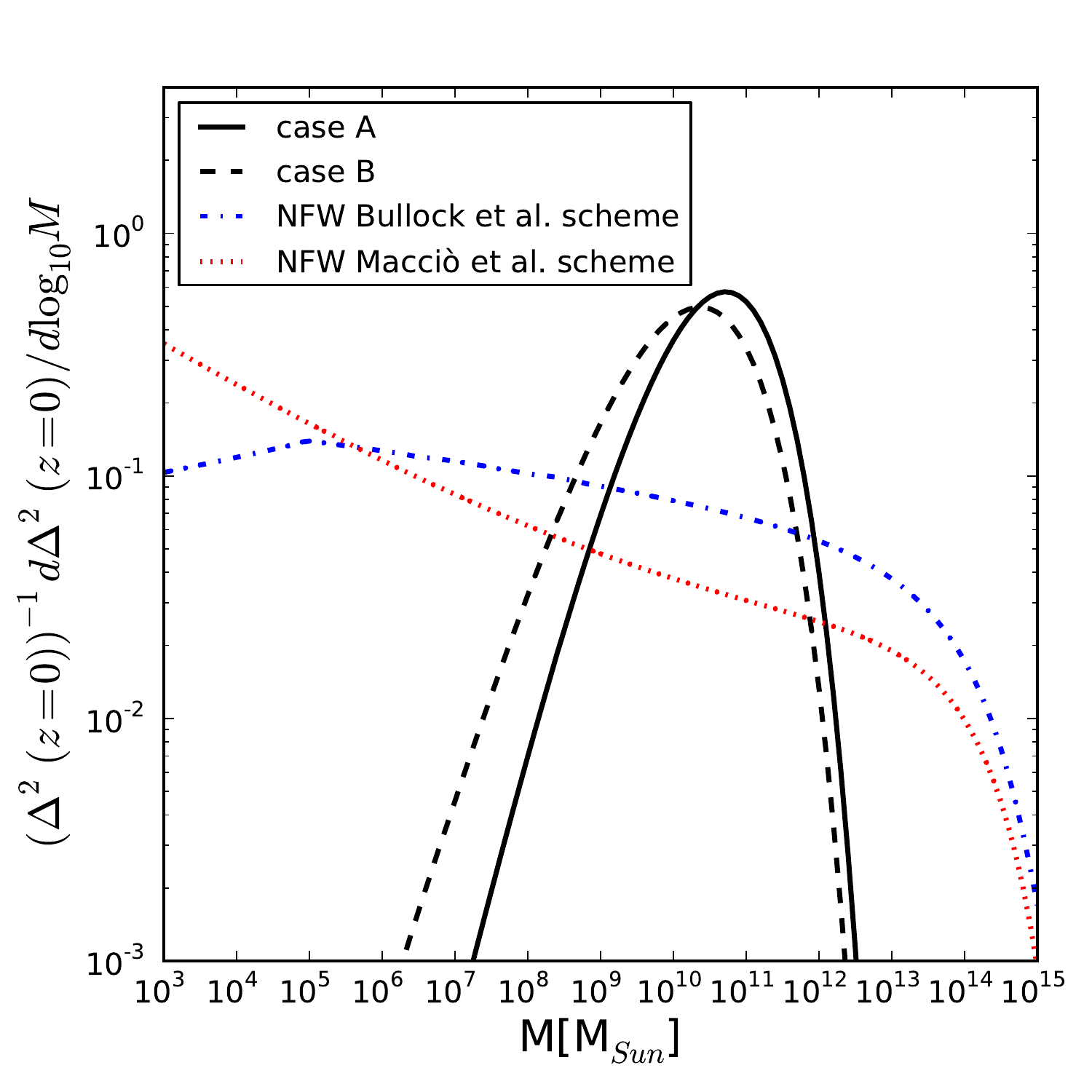} &
\includegraphics[width=0.45\linewidth,clip=]{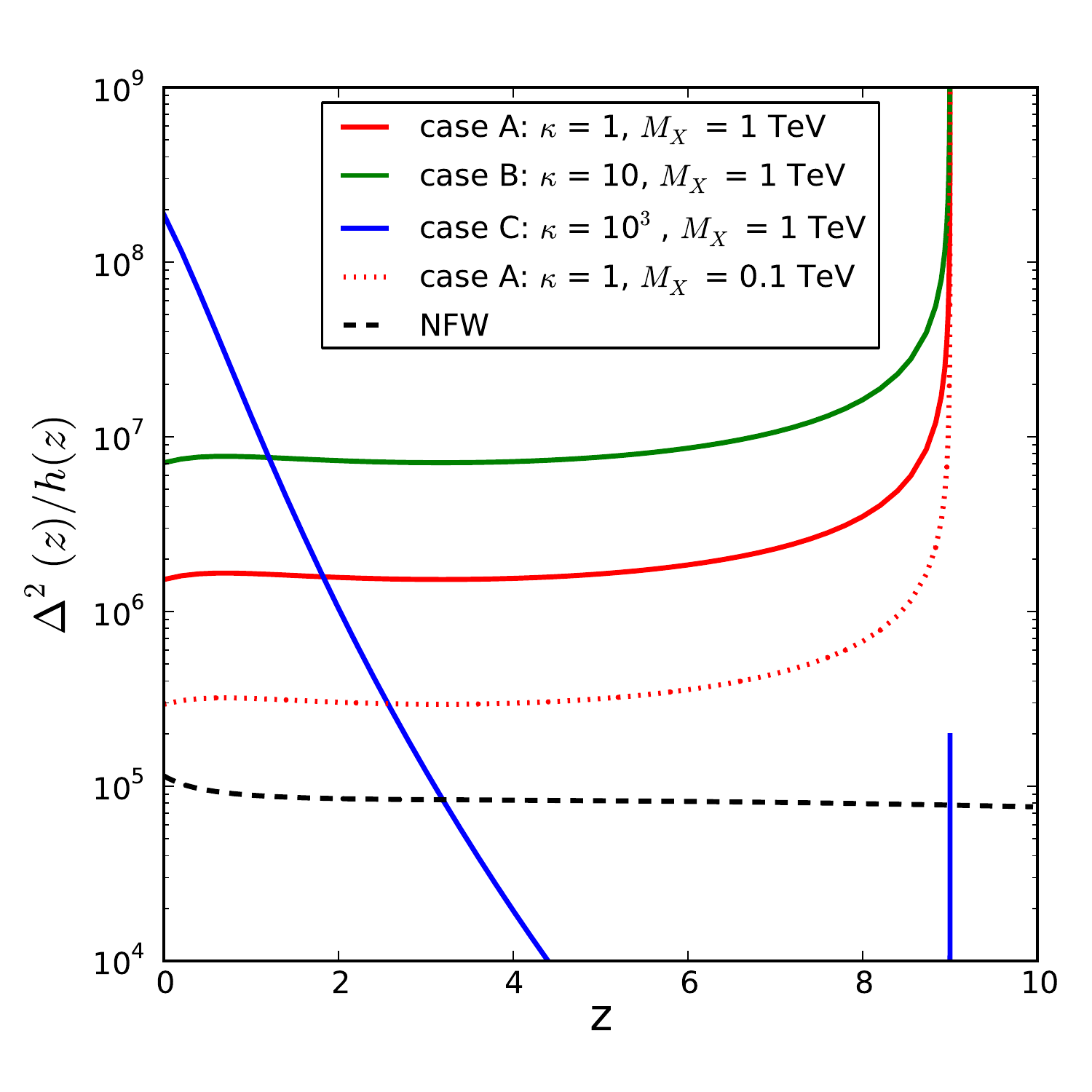}
\end{tabular}
\caption{Left: 
The distribution function $(\Delta^{2}(0))^{-1}d\Delta^2(0,M) /d \log_{10} M$ of the annihilation enhancement $\Delta^2(z=0)$ per logarithmic mass interval for case A of dark matter spikes (solid, black), case B of dark matter spikes (dashed, black), NFW dark matter halos with Bullock et al. concentration scheme \cite{Bullock:1999he} (dashed, blue), where the mean concentration is fixed to a constant below $10^5$ Solar masses (and equal to that of a $10^5$ solar mass halo) and NFW dark matter halos with the Macci\`{o} et al. concentration scheme \cite{2008MNRAS.391.1940M} (dotted, red). 
Right:
$\Delta^2(z)/h(z)$ as a function of redshift $z$ for the spikes of dark matter of masses of 1 TeV mass for cases A (solid, red), B (solid, green) and C (solid, blue) and of 0.1 TeV mass for case A (dotted, red) and $\Delta^2(z)/h(z)$ for standard NFW case of 0.1 TeV (black, dashed).}
\label{fig:DistrAndDelta_masses}
\end{figure*}

In our analysis, we consider three cases, which are summarized in Table \ref{tab:Cases}. In case A, the SMBH are formed at redshift $z_f$ and the surrounding spikes are formed at the same redshift.
The mass of each black hole is fixed to be a fraction $\alpha=10^{-3}$ of the mass of the hosting halo. We assume that the spikes are formed simultaneously at $z_f = 9$. The mass of the spike is determined by the $M_{bh}-\sigma$ relation and is a function of the mass of the hosting halo at redshift $z_f$. In case A, we take $\kappa = 1$. Therefore in case A, the redshift dependence of $\Delta^2(z)$ is carried forward only in the inner radius of the spike $r_{lim}(z)$. In case B, we define the mass of the spike by requiring $M_{sp} = 2 M_{bh}$ and use $\kappa = 10$, the rest of the model being the same as in case A.
In case C, compared to case A, we allow the black holes (and correspondingly the spikes) to grow along with the hosting halos. In eq.~(\ref{eq:deltaSq}), we assume that the mass function $\frac{dn}{dM_{bh}} (M_{bh}, z)$ depends on current redshift $z$.

In Fig. \ref{fig:DistrAndDelta_masses}, we plot the mass distribution function of the dark matter annihilation enhancement from dark matter spikes $\Delta^2(z=0)$ for cases A and B as well as for annihilations in a standard NFW profile with Bullock \cite{Bullock:1999he} and Macci\`{o} 
\cite{2008MNRAS.391.1940M} mass-concentration relations.
We note that while for dark matter halos the mass distribution function of the enhancement gets a larger contribution from less massive halos, the distribution function from the dark matter spikes is peaked around the value $10^9-10^{10}$ $M_{Sun}$ of the hosting halo.
The cut-off at higher masses is due to the exponential cut-off in the halo mass function, and the cut-off at lower masses is a sharp power-law, related to the peculiar dependence of the spike mass on the mass of the hosting halo.

In Fig. \ref{fig:DistrAndDelta_masses} (right inset), we plot $\Delta^2(z)$ for case A for masses 0.1 TeV and 1 TeV. In the same figure, we also plot $\Delta^2(z)$ for a dark matter mass 1 TeV for cases B and C and for the standard NFW case. For the NFW case, $\Delta^2(z)$ does not depend on the mass of the particle, while for case A, $\Delta^2(z)$ scales as $M^{2- 3/\gammasp}$.

{\renewcommand{\arraystretch}{1.5}
\begin{table}[h]
\centering
\begin{tabular}{c|c|c|c}
\hline
 case & spike radius definition & $\frac{dn_{bh}}{dM}(z)$ & $\kappa$ = $r_{lim}$/$r^{(0)}_{lim}$\\ 
\hline
A & $M_{bh}-\sigma$ relation & $\sim\frac{dn_{halo}}{dM}(z_f)$ & 1 \\
B & $M_{sp} = 2 M_{bh}$ relation & $\sim\frac{dn_{halo}}{dM}(z_f)$ & 10 \\
C & $M_{bh}-\sigma$ & $\sim\frac{dn_{halo}}{dM}(z)$ & 1000 \\
\hline
\end{tabular}
\caption{Summary of conventions for adopted dark matter spike models.}
\label{tab:Cases}
\end{table}

In Fig.\ref{fig:EGB_masses_and_cases} (left side), we plot the preliminary results on IGRB flux measured over 44 months by {\em Fermi} LAT  \cite{Ackermann:2012} and the extragalactic gamma-ray background for case A (solid line) and for the standard NFW case (dashed line) for DM particle masses 0.1 TeV, 1 TeV and 10 TeV, annihilating to $\tau^+\tau^-$ with the cross section of $3\times 10^{-26} \textrm{cm}^3 s^{-1}$. To obtain the annihilation spectra we used the PPPC4DMID package \cite{Cirelli:2010xx}. In the same figure we also plot the the contribution to the flux from redshifts $[0.9z_f, z_f]$ for dark matter masses 0.1 TeV, 1 TeV and 10 TeV. Since $\Delta^2(z)$ behavior at $z \to z_f$ is $(z_f - z)^{\alpha}$, where $\alpha = 3/\gammasp - 2 > -0.8$, though this mild singularity is reduced by the fact that the annihilation for each spike is cut off at few times the Schwarzschild radius, the contribution to the flux from the range of redshifts $[z_a, z_f]$ has a weak dependence on the lower limit of integration.

One observes that while for 0.1 TeV, the fluxes from dark matter spikes and NFW halos are similar in magnitude, the fluxes for dark matter spikes corresponding to dark matter particles of higher masses are higher than their NFW counterparts. In the NFW case, the flux scales with $M_X^2$, while in the case of dark matter spikes, the scaling is $M^{-3/\gammasp}_X$ due to the smaller inner radii of the spikes for higher dark matter particle masses at a given redshift. 

Due to this phenomenon, for masses greater than 100 GeV and for all cases considered and $\kappa = 1$, the extragalactic gamma-ray background flux arising from density spikes is always greater than that for the NFW case.

In Fig.\ref{fig:EGB_masses_and_cases} (right side), the extragalactic gamma-ray background is plotted for a dark matter particle mass 1 TeV for cases A,B and C and for the standard NFW case for the same annihilation channel and cross section. While it is tempting to invoke dark matter annihilations for the explanation of the extragalactic gamma-ray background, the background might also be due to a variety of astrophysical classes of sources. Among these, flat spectrum radio quasars \cite{Stecker:2010di} are the most studied, but there are also significant contributions from star-forming galaxies, radio galaxies and Bl Lacertae objects. In the same figure, we plot the spectrum of star-forming galaxies for the case when the spectrum is determined from the IR luminosity function model (dashed line), and the integrated flux of flat spectrum radio quasars (dotted line) \cite{Stecker:2010di}. 
The predicted spectrum for star-forming galaxies strongly depends on the assumptions of the model and the magnitude accordingly varies by an order of magnitude. 
It is important to note that the sum of all astrophysical contributions (star-forming galaxies, flat spectrum radio quasars, radio galaxies and Bl Lacertae objects) does not explain the diffuse background signal at all energies, and therefore one can reasonably argue that an extra component might be required \cite{Gomez-Vargas:2013cna}.

We find that adoption of the non-relativistic vs relativistic cut-off radius $r_c$, as mentioned in section \ref{sec:InnerRadius}, despite having a significant effect on the annihilation rate of a given dark spike around its formation time \cite{Sadeghian:2013laa}, has implications only at the few percent level for the diffuse gamma-ray background.
However, the variation of other parameters of our model, defining the physics of dark matter annihilating in spikes, can have an order of magnitude effect on the magnitude of the gamma-ray flux. 

Our setup differs from that of \cite{Ahn:2007ty} in several aspects, but despite these differences our results are in qualitative agreement.  
The differences include the choice of the black holes formation redshift ($z_f=9$ in our case, as opposed redshifts from $z_f=6$ to $z_f=2$), the choice of linear relation between the mass of the SMBH and the mass of the halo at the formation redshift, as opposed to the well-motivated power law relation between the mass of the SMBH and the halo at redshift zero. Moreover, in our setup we neglect possible spike disruption by stellar heating.
The assumption of later formation redshift diminishes the spikes' contribution to the gamma-ray flux, while neglecting the disruption due to stellar heating enhances it.

We find that for $\sigma v = 3\times 10^{-26} \textrm{cm}^3 s^{-1}$ and for $M_X = 100 \,\text{GeV}$ the prediction for the contribution to the IGRB flux from dark matter spikes, formed around SMBHs with $z_f = 9$ is about an order of magnitude above the contribution from NFW dark matter halo, while in \cite{Ahn:2007ty} find the same preeminence of the spikes over the NFW halos for SMBHs with $z_f = 2$.

In \cite{Horiuchi:2006de} the contribution of dark matter spikes from intermediate mass black holes have been studied and it was found that for the same case  $\sigma v =3\times 10^{-26} \,\textrm{cm}^3 s^{-1}$ and $M_X = 100 \: \text{GeV}$ depending on the scenario, the spikes' contribution can be 10 to 1000 times greater than that of the NFW halos. The factor of a thousand might be due to averaging over halo masses which resulted in assuming that every halo contains about a hundred of black holes of mass $10^5$. While the formation mechanisms of intermediate mass black holes and SMBHs are different, the boundary between the supermassive and the intermediate mass ranges might be diffuse, with $M_{bh}= 10^5$ being close to the lower range of SMBH masses.

In this paper we have considered the contribution of annihilating dark matter to the IGRB. 
While it has been widely accepted (sometimes implicitly) that annihilations from dark matter halos should contribute isotropically to the gamma-ray background, more recent studies reveal that annihilating dark matter does induce quantifiable anisotropy in the angular power spectrum \cite{Fornasa:2012gu} (primarily due to rare unresolved substructures)
and thus anisotropy measurements can be a powerful tool to disentangle astrophysical contributions from the gamma-ray background \cite{Ando:2006cr}. The role of anisotropies cannot be overstated: in the case of blazars, the anisotropy constraint turns out to be more stringent than the ones derived from the intensity alone \cite{Cuoco:2012yf}. 

In the case of dark matter density spikes, the contribution to the IGRB is expected to be even less isotropic than that of dark matter halos, due to to the fact that a narrower range of halos is expected to contribute.Hence the anisotropy might serve as a powerful tool \cite{Ando:2006cr, Ando:2013ff} to distinguish the dark matter spikes contribution from the contribution of rare sources such as blazars. It will be of interest to pursue this in a future study.

Dark matter spikes might be present not only around SMBHs but also around intermediate mass black holes at centers of dark matter subhalos \cite{Bertone:2005xz}. In this case despite the fact that the dark matter spikes would be more numerous, since intermediate mass black holes are Population III star remnants \cite{Madau:2001sc}, the extragalactic gamma-ray signal from them is expected to be smaller than that from the SMBH dark matter spikes \cite{Horiuchi:2006de}.

The prompt products of dark matter annihilations are likely to interact in the interstellar and intergalactic medium to produce a flux of secondary emission at lower energies. For example, the electrons and positrons resulting from dark matter annihilations interact with the cosmic microwave background via inverse Compton scattering and might provide constraints in some scenarios that are more competitive than those from the primary annihilation products \cite{Profumo:2009uf, Belikov:2009cx}.

\begin{figure*}[h]
\centering
\begin{tabular}{ccc}
\includegraphics[width=0.45\linewidth,clip=]{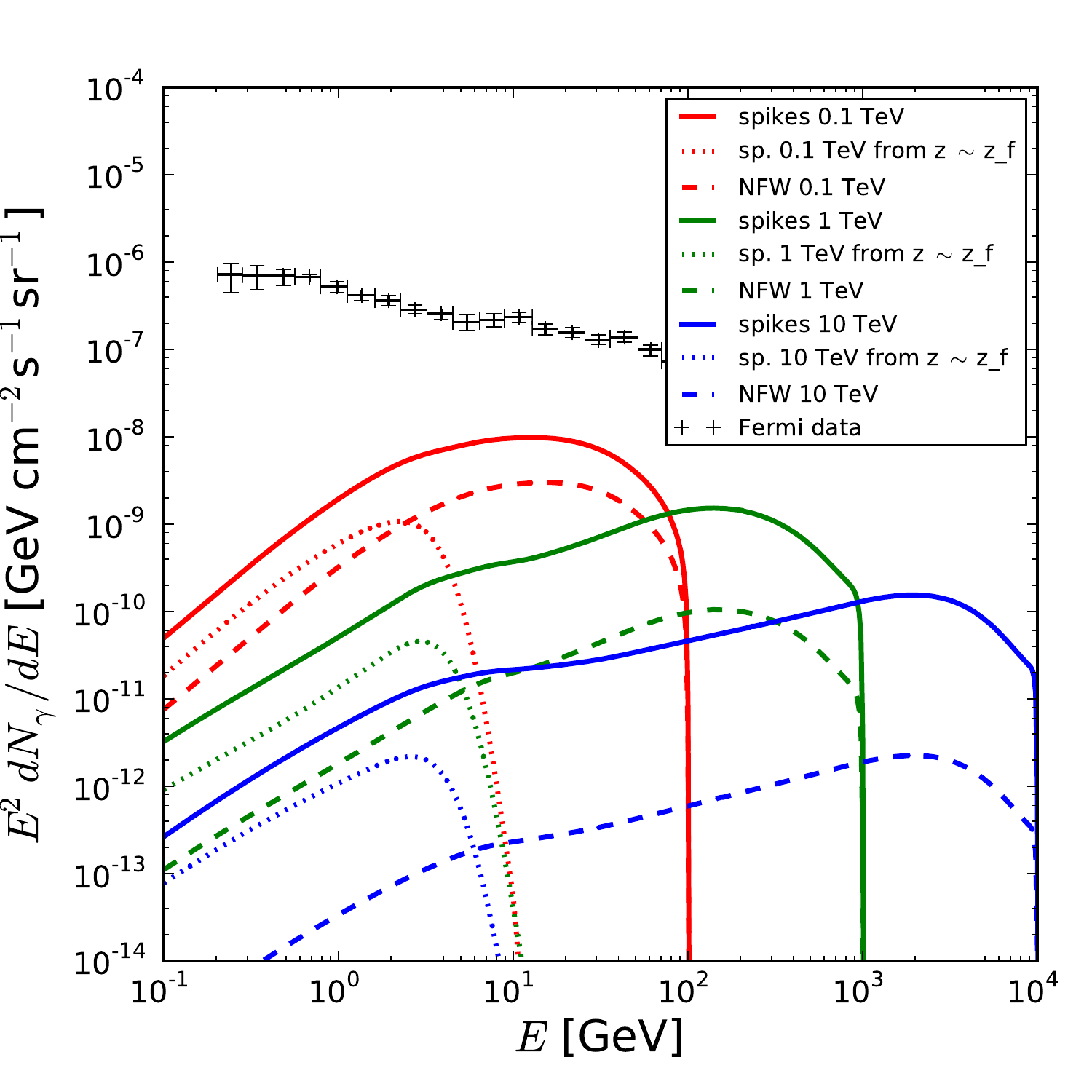} &
\includegraphics[width=0.45\linewidth,clip=]{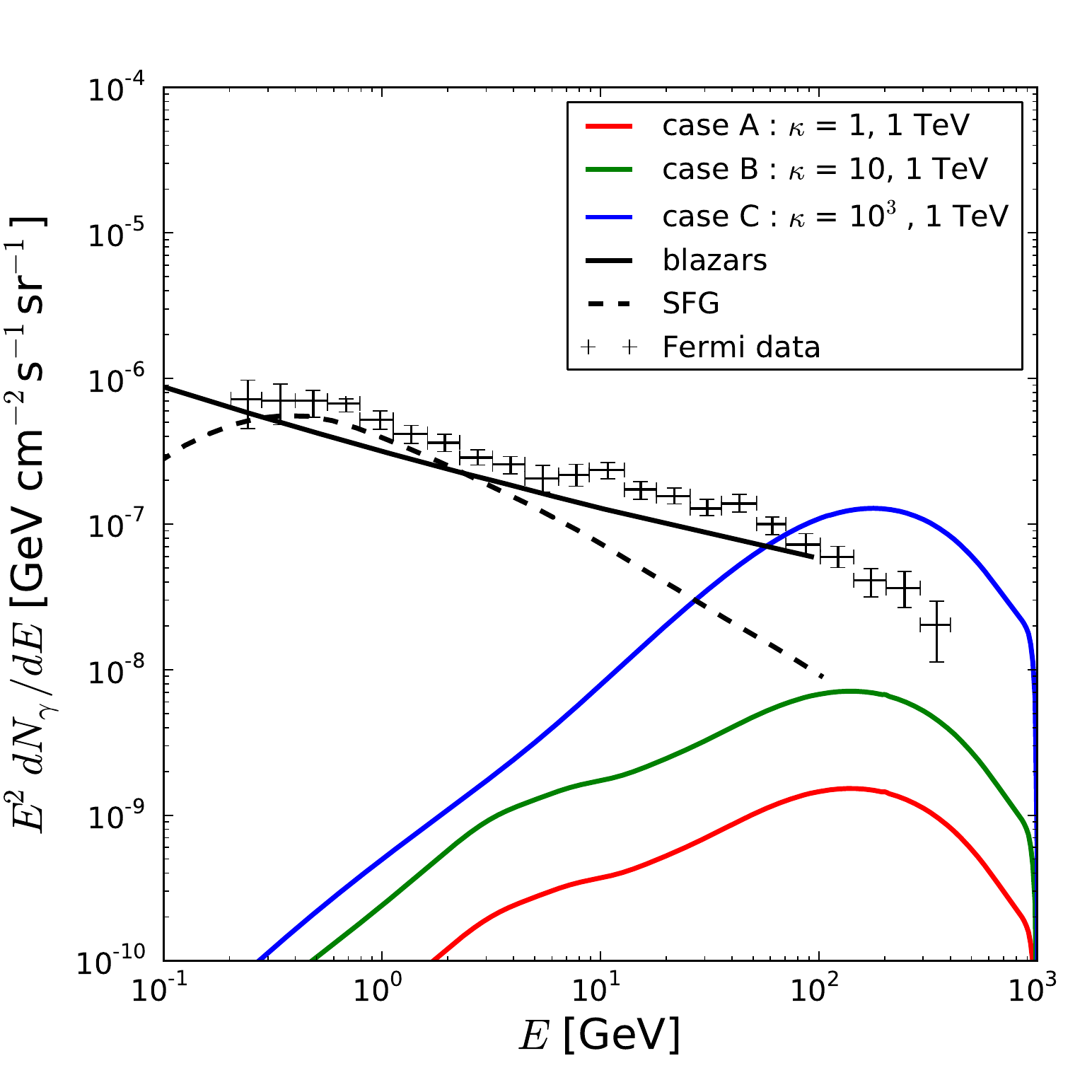}
\end{tabular}
\caption{Left: The IGRB flux measured by {\em Fermi} LAT and the predicted flux for spikes of dark matter of masses of 100 GeV, 1 TeV and 10 TeV (red, green, blue; solid lines), the flux contribution from $z \simeq z_f$ for spikes of dark matter of masses of 100 GeV, 1 TeV and 10 TeV (red, green, blue; dashes lines) and the standard NFW flux for dark matter of masses of 100 GeV, 1 TeV and 10 TeV (red, green, blue; dotted lines). All the fluxes are calculated to $\tau^+\tau^-$ annihilation channels.
Right: 
The IGRB flux measured by {\em Fermi} LAT and the predicted flux for dark matter spikes cases A, B and C (red, green, blue) for dark matter of mass 1 TeV. All the fluxes are calculated to $\tau^+\tau^-$ annihilation channels. The predicted flux for dark matter (1 TeV) spikes case A (red), flat spectrum radio quasars (green) and star-forming galaxies (blue, dashed) \cite{Stecker:2010di}.}
\label{fig:EGB_masses_and_cases}
\end{figure*}

\section{Conclusions} 
\label{sec:concl}

We have considered the contribution of dark matter annihilations in density spikes around central supermassive black holes in dark halos to the isotropic extragalactic gamma-ray background. We introduced several models for dark matter spikes, relating the mass of the spike to the mass of the hosting halo and the spike mass function to the halo mass function. We introduced a parameter $\kappa$ to describe in an {\it ad hoc} way the reduction of annihilation rates due to the disruption of dark matter spikes by astrophysical processes such as merging or stellar heating. We have verified that lowering the minimum limiting radius of the spike to two Schwarzschild radii as follows from the relativistic calculation does not have a significant effect on the gamma-ray background signal. We have demonstrated that for $M_X$ = 100 GeV and $\kappa = 1$, the contribution of dark matter spikes to the annihilating dark matter signal considerably exceeds that from dark matter halos, in some cases by orders of magnitude. Astrophysical models, such as blazars, for the high energy component of the extragalactic gamma-ray background should be distinguishable from our dark matter annihilation model by both spectral and anisotropy signatures.

\begin{acknowledgments}
This research has been supported at IAP by the ERC project 267117 (DARK) hosted by Universit\'e Pierre et Marie Curie - Paris 6 and at JHU by NSF grant OIA-1124403. We are grateful to an anonymous referee for useful comments and suggestions. We thank the Kavli Institute for Theoretical Physics and the Kavli Institute for Cosmological Physics for hospitality during various stages of the project.
\end{acknowledgments}

\bibliography{bibl}

\end{document}